\documentclass[aps,pre,reprint,groupedaddress,showpacs]{revtex4-1}
\usepackage{graphicx}
\usepackage{epstopdf}

\usepackage{amssymb}
\usepackage{amsmath}
\usepackage{psfrag}
\usepackage{pstool}

\begin{document}


\title{The solidification of a disk-shaped crystal from a weakly supercooled binary melt}


\author{David W. \surname{Rees Jones}} 
\email[]{David.ReesJones@physics.ox.ac.uk}

\author{Andrew J. \surname{Wells}}

\affiliation{Atmospheric, Oceanic and Planetary Physics, Department of Physics, University of Oxford, Clarendon Laboratory, Parks Road, Oxford, OX1 3PU, UK.}



\begin{abstract}
The physics of ice crystal growth from the liquid phase, especially in the presence of salt, has received much less attention than the growth of snow crystals from the vapour phase. The growth of so-called frazil ice by solidification of a supercooled aqueous salt solution is consistent with crystal growth in the basal plane being limited by the diffusive removal of the latent heat of solidification from the solid--liquid interface, while being limited by attachment kinetics in the perpendicular direction. This leads to the formation of approximately disk-shaped crystals with a low aspect ratio of thickness compared to radius, because radial growth is much faster than axial growth.  We calculate numerically how fast disk-shaped crystals grow in both pure and binary melts, accounting for the comparatively slow axial growth, the effect of dissolved solute in the fluid phase and the difference in thermal properties between solid and fluid phases. We identify the main physical mechanisms that control crystal growth and show that the diffusive removal of both the latent heat released and also the salt rejected at the growing interface are significant.  Our calculations demonstrate that certain previous parameterizations, based on scaling arguments, substantially underestimate crystal growth rates by a factor of order 10-100 for low aspect ratio disks, and provide a parameterization for use in models of ice crystal growth in environmental settings. 
\end{abstract}


\maketitle



\section{Introduction} \label{sec:disc-intro}
Ice is a particularly rich example of crystallization, with a wide range of crystal shapes formed depending on the environmental conditions \citep{Libbrecht05}. It is also environmentally significant: it forms from the vapour phase in clouds, leading to snow and sleet, and from the liquid phase in rivers and oceans. We study so-called frazil-ice formation from the liquid phase in the environmentally relevant limit of weak supercooling, because this has received comparatively little attention \citep{Pruppacher97}. It also has key applications, both in industrial settings where frazil ice can block the water inlets from rivers and lakes \citep{Daly91}, and in geophysical settings where frazil ice forms under floating ice shelves and in open areas of the polar oceans called leads and polynyas \citep{MartinReview81}. 

Frazil ice consists of individual crystals as a particulate suspension in a supercooled liquid from which the ice grows. This liquid could be freshwater, such as when frazil forms in rivers, or saltwater, such as when frazil forms in the ocean.  We study crystal growth from a binary alloy as a simple proxy for saltwater. Crystal growth from a binary alloy has been studied in a variety of geometries, most extensively for spherical and axially-symmetric cylindrical crystals, where morphological instability has been shown to be significant leading to dendritic growth \citep{Mullins63,Coriell65}. 

Crystallization is often an inherently anisotropic process, and macroscopic anisotropy can arise from crystalline anisotropy. Anisotropic surface energy is responsible for the so-called `equilibrium Wulff shape' and anisotropic kinetic attachment is responsible for the `kinetic Wulff shape,' as reviewed by \citet{Sekerka05}. Both these physical effects play an important role in the faceted growth of snow ice from the vapour phase, as shown by the numerical study of \citep{Barrett12}. However, a roughening transition can occur for solidification with sufficiently small supercooling, leading to the rapid growth of certain faces with finite curvature rather than planar faceted faces \citep{Dash06}.  For ice growth from vapour, the roughening transition occurs in prism faces for temperatures above $-2^\circ$C \citep{Elbaum91}, whilst for ice growth from the liquid phase, the roughening transition  occurs above $-16^\circ$C  \citep{Maruyama05}. Basal facets have been observed to persist until the melting temperature in both cases \citep{Elbaum91,Maruyama05}. We here consider crystal growth in the liquid phase, at temperatures above the roughening transition for faces that grow in the basal plane. We focus on solidification controlled by the long range diffusive transport of heat and salt, and how this couples with anisotropic kinetic attachment in determining bulk crystal growth from the melt. Frazil ice is observed to form axisymmetric disk-shaped crystals, at least for fairly weak supercooling \citep{Martin81,Ghobrial13,McFarlane14}. Slow attachment kinetics limit growth perpendicular to the basal plane of the crystal while growth in the basal plane is limited by diffusion \citep{Daly84,Pruppacher97}. 

Previous studies make various approximations in order to determine the growth rate of a disk-shaped crystal in a pure melt. Some proceed by the well known `electrostatic analogy' between ice growth limited by thermal diffusion and electrostatic capacitance \citep{McDonald63}. To give an example, \citet{Mason58} uses this method to estimate the mass growth rate of a disk from the vapour phase. In this analogy, temperature is analogous to electrostatic potential, and the crystal growth rate is proportional to the capacitance of a perfect conductor, the surface of which will have a constant potential. Thus, knowing the capacitance of a thin disk and assuming its thickness evolves slowly, the radial growth rate can be estimated. The analogy assumes that the disk is perfectly conducting (which is a good approximation for ice growth from the vapour phase but not the liquid phase) and infinitesimally thin. We will therefore assess whether the limits of perfectly conducting and infinitely thin can be taken independently. 

Other studies note that growth controlled by the diffusive removal of latent heat depends on the ratio of the rate of latent heat release to the rate of thermal transfer away from the interface \citep{Daly84,Holland07}. These rates can in principle be estimated to within an undetermined dimensionless prefactor using scaling analysis. The success of a scaling approach relies on identifying the most appropriate physical scales.

A more detailed study was made by \citet{Fujioka74}, who make the mathematical simplification of assuming that the material properties of the phases are equal and that the growth was purely radial. The authors found a separable solution for the temperature field subject to diffusive heat transfer. This model has been used by \citet{Yokoyama09} to explain the experimental observations of \citet{Shimada97} of the growth and stability of ice crystals from a pure melt. 

The goal of this paper is to quantify the leading order factors controlling growth of disk-shaped crystals from a binary melt, accounting for the impacts of diffusive heat and salt transfer, and different material properties in a setting with dominant radial growth. We also make calculations with an approximate model of axial growth.  None of these generalisations can be tackled using the methodology of \citet{Fujioka74}. To isolate the separate physical effects, we build the complexity of our model in stages. We first consider the effect of the geometric shape of the crystals and the different material properties of the phases by considering the growth of disk-shaped crystals in a pure melt (section \ref{sec:pure}). We then consider the effect of axial growth (section \ref{sec:axial}) and the effect of salt by considering a binary alloy (section \ref{sec:binary}). Finally, we discuss the physical significance and implications of our results (section \ref{sec:discussion}).

\section{Growth into a pure melt} \label{sec:pure}

\subsection{Governing equations}
\begin{figure}
\includegraphics[width=8.6cm]{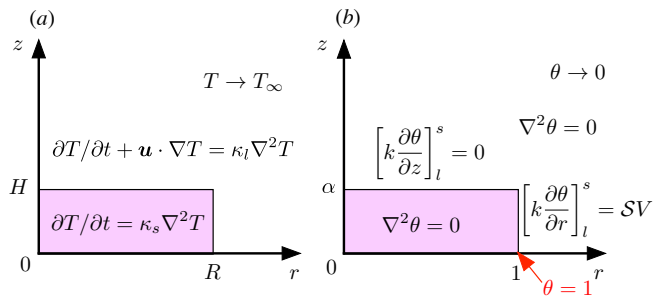}
\caption{(\textit{a}) Dimensional problem in cylindrical  polar coordinates showing the heat equation inside and outside a disk-shaped crystal of radius $R$ and half-thickness $H$. (\textit{b}) dimensionless, quasi-steady problem with $\boldsymbol{u}=0$ and simplified boundary conditions for purely radial growth explained in the text. Note that $\theta=1$ is applied at $(r=1, z=0) $.} \label{fig:diagram}
\end{figure}

We first introduce the equations and boundary conditions used to determine the growth of an isolated crystal into a pure melt. Consider an isolated  axisymmetric disk-shaped crystal, as shown in figure \ref{fig:diagram}, of radius $R$ and half-thickness $H$, such that the aspect ratio $\alpha=H/R$, which we expect to be small. To aid progress with modelling, we make the simplifying assumption that crystal growth maintains the disk like geometry observed in experiments \citep{Martin81,Ghobrial13,McFarlane14} with uniform growth rates across each individual crystal face. 
This is a reasonable approximation for radial growth of the thin disk edges provided that the disk remains morphologically stable (discussed further in section \ref{sec:limitations}). Macroscopically uniform axial growth is consistent with an activated process, where growth is limited by the difficulty in nucleating a new layer of molecules somewhere on the face. This is followed by more rapid spreading of the layer following nucleation such that the interface remains macroscopically flat \citep{Wettlaufer94}. Alternatively, quasi-planar faces have been maintained in numerical models with anisotropic surface energy and anisotropic attachment kinetics \citep{Barrett12}, where the variation in interfacial temperature across the crystal face (arising from the so-called Berg effect \citep{Berg38}) can be offset by weak interfacial curvature. Our treatment of axial growth is discussed further in section \ref{sec:axial}. We introduce cylindrical polar coordinates $(r,\phi,z)$, where the $z$-axis is perpendicular to the basal plane and the origin is the centre of the crystal. The temperature $T$ obeys the heat equation 
\begin{align}
\rho_s c_s \frac{\partial T}{\partial t} &= k_s \nabla^{2} T, \qquad \boldsymbol{x}\, \in \, D, \label{eq:heatD1} \\ 
\rho_l c_l \left(\frac{\partial T}{\partial t} +\boldsymbol{u} \cdot \nabla T \right) &= k_l \nabla^{2} T, \qquad \boldsymbol{x}\, \notin \, D, \label{eq:heatL1}
\end{align}
where $D$ denotes the disk-shaped crystal. The density $\rho$, specific heat capacity $c$, and thermal conductivity $k$ take constant values in each phase, whether solid (subscript $s$) or liquid (subscript $l$), $\boldsymbol{u}$ is the fluid velocity, and $t$ is time. The thermal diffusivity $\kappa=k/\rho c$. We assume that $T$ approaches a uniform temperature $T_\infty$ far from the crystal. 

We impose a regularity condition at $r=0$ and a symmetry boundary condition at the mid-plane of the disk ($z=0$) so that we may restrict attention to $z\geq0$. At the boundary $\partial D$ of the disk, suitable boundary conditions result from heat conservation and a kinetic condition of thermodynamic disequilibrium, respectively
\begin{align}
\rho_s L V_{\mathrm{dim.}} &=  \left[k\frac{\partial T}{\partial n}\right]^s_l, \qquad \boldsymbol{x}\, \in \, \partial D, \label{eq:Stefan} \\
V_{\mathrm{dim.}} &= G\left(\boldsymbol{n},T_m-T_i\right), \qquad \boldsymbol{x}\, \in \, \partial D. \label{eq:kinetic} 
\end{align}
The temperature is continuous across the interface and equals $T_i$. A discontinuity in the heat flux at the interface is associated with the latent heat of fusion $L$ associated with crystal growth at a  velocity $V_{\mathrm{dim.}}$ normal to the interface. The normal growth is uniform across each crystal face because we assumed that the crystal remains disk-shaped. In the second equation for attachment kinetics, the function $G$ depends on the normal direction to the interface $\boldsymbol{n}$ and the difference between the equilibrium melting temperature and the interfacial temperature. The exact form used is not crucial in what follows in this section where we consider the limit of negligible axial growth, but does impact predictions of weak axial growth in section~\ref{sec:axial}, where we discuss the physical significance of $G$.

\subsection{Reduced, non-dimensional model equations for purely radial growth}
We make a quasi-steady approximation in which we neglect the explicit time-dependence of the problem in the heat equations (\ref{eq:heatD1}, \ref{eq:heatL1}), and justify this approximation a posteriori in section \ref{sec:QS}. We neglect externally driven fluid flow and buoyancy-driven flow, and note that the expansion flow (caused by the density of ice being lower than that of water) can be neglected consistently with our quasi-steady approximation, so $\boldsymbol{u}=0$. We non-dimensionalize lengths with respect to the instantaneous disk radius $R$, time with the thermal diffusion timescale in the liquid $R^2/ \kappa_l$, and velocities with $\kappa_l/R$.  We define a dimensionless temperature $\theta=(T-T_\infty)/ \Delta T_\infty$, where the supercooling is  $\Delta T_\infty=T_m-T_\infty$. 

Growth is much slower in the direction perpendicular to the crystal basal plane because it is kinetically unfavourable. Thus, in this section, we introduce a strong form of anisotropy into the model in a simple limiting form by taking $G=0$ at the top of the crystal $z=\alpha$ and $G\rightarrow\infty$ at the edge $r=1$. This limit prevents axial growth and leaves the temperature $T_i$ at $z=\alpha$ unconstrained, so it can depart from the melting temperature $T_m$. By contrast, $T_i=T_m$ at $r=1$. We consider the effect of slow axial growth $G>0$ later in section \ref{sec:axial}. The simplified boundary conditions are
\begin{equation} \label{eq:bctop}
\overline{k} \left. \frac{\partial \theta}{\partial z}\right|_{z=\alpha^{-}}= \left. \frac{\partial \theta}{\partial z}\right|_{z=\alpha^+} ,     
\end{equation}
on the top the crystal $(0\leq r \leq1)$, and
\begin{align}
\mathcal{S} V &=\overline{k} \left. \frac{\partial \theta}{\partial r}\right|_{r=1^{-}}- \left. \frac{\partial \theta}{\partial r}\right|_{r=1^+},    \label{eq:bcDTDR} \\
1 &=  \left.\theta \right|_{r=1},   \label{eq:bcT} 
\end{align}
on the edge $ (0\leq z\leq \alpha)$. 

In this quasi-steady limit, there are only three dimensionless parameters in the problem: the aspect ratio $\alpha$, the conductivity ratio  $\overline{k}=k_s/k_l$ and the Stefan number
\begin{equation}
\mathcal{S}=\frac{\rho_s L}{\rho_l c_l \Delta T_\infty}.
\end{equation}
Thus the problem reduces to solving Laplace's equation $\nabla^2 \theta=0$ in the whole domain including the disk subject to the boundary conditions (\ref{eq:bctop}--\ref{eq:bcT}). Note that $V$ is determined implicitly as part of the solution. In particular, we calculate a rescaled growth rate
\begin{equation} \label{eq:fdef} 
f(\alpha,\overline{k})=\mathcal{S}V \alpha,
\end{equation}
which is a function of the aspect ratio $\alpha$ and conductivity ratio $\overline{k}$ alone. Given that $R=1$ in our non-dimensionalization, $f$ is proportional to the growth rate multiplied by the area of the growing surface.

\begin{figure}
   \centering    \includegraphics{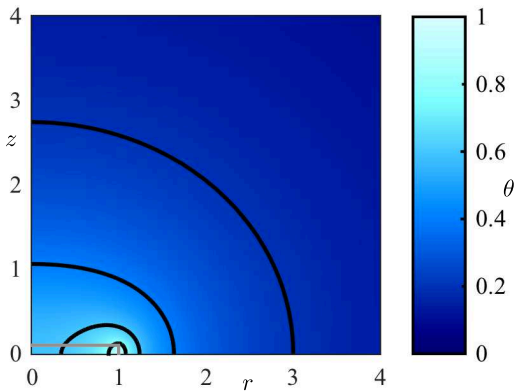}
   \caption{Example of heat transfer when $\alpha=0.1$, $\overline{k}=1$, with  contours $\theta=0.2, 0.4, 0.6, 0.8$ shown. Note that the thermal boundary layer scales with the disk radius rather than the disk thickness. The crystal boundary is shown as a grey line.}  \label{fig:EG-heat}
   \end{figure}

The boundary conditions contain a subtlety in that equations \eqref{eq:bcDTDR} and \eqref{eq:bcT} are formally inconsistent. This is evident upon studying solutions to Laplace's equation near the `corner' between the top and edge of the disk \citep[\textit{cf}.][]{Jackson99}. The inconsistency arises from the simplifying assumption that the crystal remains perfectly disk shaped. In reality, we might expect the crystal shape to evolve via non-uniform growth rates and the regularising impact of surface energy described by the Gibbs-Thomson effect \citep[\textit{e.g.}][]{Davis01}, with the freezing temperature modified by curvature generated near the `corner' over a length comparable to the capillary length scale. Our primary interest is in leading order scalings for the macroscopic relief of supercooling and the volumetric rate of ice growth, rather than the detailed microstructure of the crystal edges which will depend on the poorly constrained orientation-dependence of the anisotropic surface energy and growth rate. Hence we simplify the analysis and neglect these deviations from disk-shaped geometry. We follow \citet{Fujioka74}, imposing \eqref{eq:bcDTDR}  on $0\leq z\leq \alpha$ but only imposing \eqref{eq:bcT} at $z=0$. We will see later that the dominant thermal gradients driving ice growth scale with the crystal radius $R$ rather than the disk half-thickness $H$, and thus we expect the detailed geometry near the disk edges to have a relatively small influence on macroscopic ice growth rates for thin discs with $\alpha=H/R\ll 1$. We have also tested the converse approach of applying equation  \eqref{eq:bcT} on  $0\leq z\leq \alpha$ but only imposing \eqref{eq:bcDTDR} in an integral sense. The difference is negligible for $\alpha \ll 1$, suggesting that the detailed micro-evolution of crystal shape does not play a very significant role in radial disk growth from the liquid phase. We discuss our numerical method in appendix \ref{app:numerics}, and show a typical solution in figure \ref{fig:EG-heat}.

\subsection{Comparison with previous models}
The function $f$ represents a crystal growth rate scaled with the Stefan number. To aid comparison, we define equivalent growth rate functions below based on previous published models. 

The electrostatic analogy model of \citet{Mason58} gives
\begin{equation}
f_M=\frac{2}{\pi}.
 \end{equation}
This is consistent with our function $f(\alpha,\overline{k}) \rightarrow 2/\pi$ as $\overline{k}\rightarrow \infty$ and $\alpha \rightarrow 0$. Note that the Mason model has no dependence on aspect ratio. 

A commonly applied scaling argument of \citet{Daly84}, gives
\begin{align}
f_D &=\alpha \sqrt{2/(1+2\alpha)}, \nonumber \\
&\sim \alpha\sqrt{2} \quad (\alpha \rightarrow 0),
\end{align}
which has a strong dependence on aspect ratio and  predicts much smaller growth for thin disks than the Mason model.

For equal thermal conductivities $(\overline{k}=1)$, the model of \citet{Fujioka74} gives
\begin{equation}
 f_{FS}= \frac{\pi \alpha}{q_0(\alpha)},
\end{equation}
where 
\begin{align} 
q_0(\alpha) &=2 \int_0^\infty \frac{\sin (\alpha x)}{x} I_0(x) K_0(x) \, dx, \nonumber \\
 & \sim \alpha\left[1+3 \ln(2) - \ln(\alpha)\right]  \quad (\alpha \rightarrow 0), \label{eq:toroidal}
\end{align} 
is the toroidal integral of order zero in which $I_0$ and $K_0$ are the modified Bessel functions of order zero. This implies that the growth rate has only a weak, logarithmic dependence on aspect ratio.

\begin{figure} 
\centering    \includegraphics[width=8.6cm]{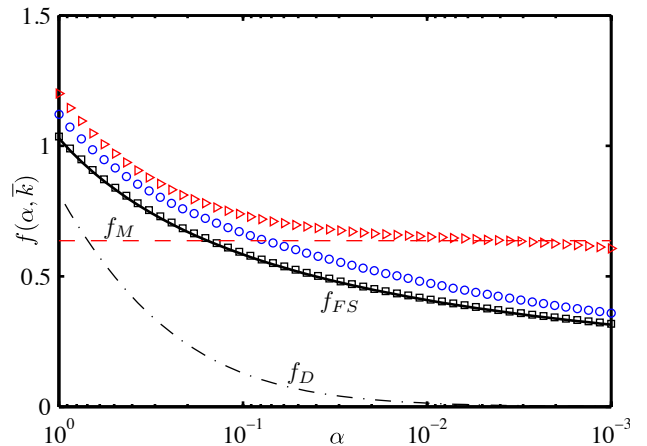}
\caption{Heat transfer factor $f(\alpha,\overline{k})$ in equation \eqref{eq:fdef}. Symbols show our numerical results for $\overline{k}=1$ (black squares), $\overline{k}=4$ (blue circles) relevant to a water-ice system, and $\overline{k}=1000$ (red triangles). Note that the $\alpha$ axis is reversed, motivated by the aspect ratio decreasing over time for a radially growing crystal. The red horizontal dashed line shows the \citet{Mason58} model, the solid curve shows the \citet{Fujioka74} model and the much lower dot-dashed curve shows the \citet{Daly84} model.} \label{fig:heat-comp}
\end{figure}

We present our own findings in comparison to these previous models in figure \ref{fig:heat-comp}. Note that our benchmarked numerical results for $\overline{k}=1$ agree with \citet{Fujioka74}, and   approach the scaling of \citet{Mason58} at high $\overline{k}$ and small $\alpha$. Crucially, there is a significant difference compared to the scaling of \citet{Daly84}. This difference is strongly linked to structure of the heat transport, a significant proportion of which occurs through the flat top and base of the disk. An example of the corresponding temperature field is shown in figure \ref{fig:EG-heat}. We return to this issue in section \ref{sec:scaling}.  We give approximate fits to our numerical calculations for practical use in appendix \ref{app:formulae}. 

 \subsection{Dependence on thermal conductivity ratio}
\begin{figure} 
\centering      \includegraphics[width=8.6cm]{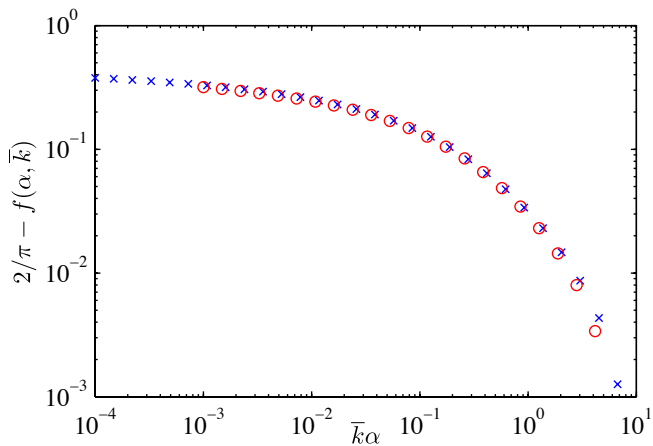}
\caption{The heat transfer factor (red circles $\alpha=10^{-3}$, blue crosses $\alpha=10^{-4}$) approaches the \citet{Mason58} model $f_M=2/\pi$ at high $\overline{k}\alpha$.}  \label{fig:heat-depk}
\end{figure}

Interestingly, even quite large changes in the thermal conductivity ratio have only a modest effect on the radial growth of crystals, with $f(\alpha,\overline{k})$ changing by less than factor of 2 as $\overline{k}$ varies over 3 orders of magnitude (figure \ref{fig:heat-comp}).  Whilst a higher solid phase conductivity transports latent heat away from the interface more efficiently, the heat flux through the solid depends on the product $\alpha\overline{k}$, which is typically small, and the heat must in any case be transported away from the disk. The \citet{Mason58} model corresponds to large $\alpha\overline{k}$, and so represents a limit which is inappropriate for frazil-ice growth in the ocean, but much more appropriate for ice formation in clouds (its original purpose) because ice is very much more thermally conductive than air. Thus the physical processes controlling ice growth differ markedly between growth from the vapour and the liquid phase. Calculations at very high values of  $\overline{k}$, shown in figure \ref{fig:heat-depk}, demonstrate  that the \citet{Mason58} model does indeed obtain the correct limiting behaviour, provided $\overline{k} \alpha \geq O(1)$.
 
\subsection{Validity of the quasi-steady approximation} \label{sec:QS}
The quasi-steady approximation is generally taken to hold provided the Stefan number $\mathcal{S}\gg1$  \citep{Mullins63}. However, while this standard requirement is appropriate for the growth of a spherical crystal, it must modified for the growth of a disk crystal. In particular, we may neglect the explicit time dependence in equation \eqref{eq:heatL1} if $V \ll 1$. Thus using equation \eqref{eq:fdef} we firstly require $\mathcal{S} \alpha\gg1$, given that $f=O(1)$ throughout the parameter range of interest. This is another reminder of the differences that arise from the geometry of crystal growth. Second, the dimensionless strength of the expansion flow (induced by the density of ice being lower than that of water) is negligible provided $\mathcal{S} \rho_l/(\rho_l-\rho_s) \gg1$, because the induced flow is proportional to solidification rate and the density difference. Third, we may neglect the time dependence in the heat equation for the solid phase \eqref{eq:heatD1} provided $\mathcal{S} \alpha\kappa_s/\kappa_l\gg1$. For ice--water disk crystals, these latter requirements are satisfied automatically provided  $\mathcal{S} \alpha\gg1$. 

\section{Non-zero axial growth} \label{sec:axial}
We investigate the potential effect of axial growth on the overall growth characteristics of the crystal by allowing a non-zero kinetic attachment coefficient for axial growth. In the previous section, we introduced an extreme, limiting form of anisotropy into the model by requiring that the disk remained at constant thickness through imposing a kinetic attachment coefficient $G=0$ on the top of the disk. In our non-dimensionalization of equations \eqref{eq:Stefan} and~\eqref{eq:kinetic}, the dimensionless kinetic coefficient is
\begin{equation}
\frac{\rho_sL R }{k_l \Delta T} \, G\left(\boldsymbol{n},T_m-T_i \right),
\end{equation}
which  may become $O(1)$  as the crystal radius increases. 
 
\begin{figure*} 
\centering      \includegraphics[width=17.2cm]{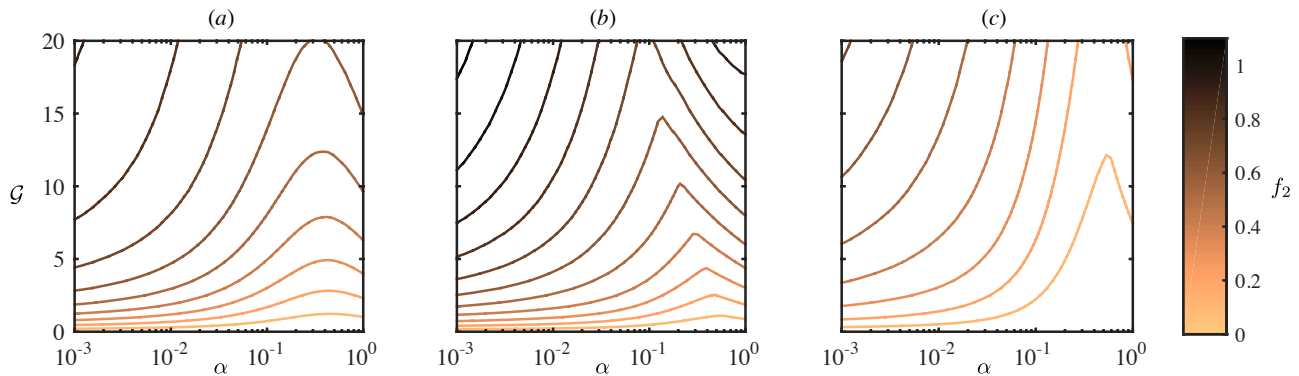}
\caption{Contours of the axial growth rate $f_2$ for growth laws given by equations (\ref{eq:ax1}--\ref{eq:ax3}) in (\textit{a}--\textit{c}) respectively. The contours have an equal spacing of 0.1. Throughout we take $\overline{k}=4$.} \label{fig:axial-laws}
\end{figure*}

The precise form of $G$ in the axial direction is under constrained, but we here investigate several illustrative cases. A simple first approach is to assume that the attachment coefficient is linearly related to the supercooling at the interface \citep{Chernov74}, $G=G_1(T_m-T_i)$, and also that this temperature difference should be averaged over the face of the disk, consistent with assuming disk-shaped growth. A second alternative is to assume that growth is determined by the maximum, rather than the average, supercooling \citep{Yokoyama09}. Furthermore,   \citet{Yokoyama09} suggest using a quadratic dependence on the maximum supercooling $G=G_2(T_m-T_i)^2$ (corresponding to a screw dislocation) on the basis of experimental observations, as discussed in \citet{Yokoyama00}. An alternatively proposed \citep{Yokoyama00} dependence of $G$ on the tenth power of the supercooling (corresponding to two dimensional nucleation) results in extremely slow growth for the small supercooling that we consider, so the results should be well approximated by the limit of negligible axial growth considered in the previous section. These mechanisms of crystal growth allow the interface to remain macroscopically flat in spite of the variation in $T_m-T_i$ across the face. We define a rescaled axial growth rate $f_2(\alpha,\overline{k},\mathcal{G})=SW$, where $W$ is the dimensionless axial growth rate. The function $f_2$ is proportional to the product of the growth rate and the growing surface area. In this section, we illustrate results relevant to frazil-ice formation from liquid water, so fix $\overline{k}=4$. These three alternatives discussed above give growth rates, in dimensionless form, 
\begin{align}
f_2 &=\mathcal{G}\,  \int_0^1 (1-\theta_i) 2r \, dr, \label{eq:ax1} \\
f_2 &=\mathcal{G} \max(1-\theta_i), \label{eq:ax2} \\
f_2 &=\mathcal{G} \max(1-\theta_i)^2, \label{eq:ax3}
\end{align}
where $\mathcal{G}=G_1 \rho_sL R / k_l $ for the linear laws (\ref{eq:ax1}--\ref{eq:ax2}) and $\mathcal{G}=G_2 \Delta T \rho_sL R / k_l $ for the quadratic law \eqref{eq:ax3}. There are some differences between the resulting axial growth rates for each law as shown in figure \ref{fig:axial-laws}. Growth based on the maximum supercooling \eqref{eq:ax2} is necessarily faster than that based on the average \eqref{eq:ax1}, and does not vary smoothly with $\alpha$ and $\mathcal{G}$, because the position of the maximum supercooling jumps. For small $\alpha$, the maximum supercooling is  always in the centre of the disk, but begins to move out as $\alpha$ increases, before jumping to the edge of the disk when  $\alpha  \approx 0.1$. The quadratic law  \eqref{eq:ax3} is qualitatively similar to \eqref{eq:ax2}, but gives slower growth.

Axial growth is coupled nonlinearly to radial growth, through latent heat release associated with axial growth. Despite some quantitative differences in axial growth illustrated in figure \ref{fig:axial-laws}, the differences between the 3 growth laws do not change the qualitative effect on radial growth. Therefore, we focus on law \eqref{eq:ax1} as a representative example. The rescaled radial growth function $f_1(\alpha,\overline{k},\mathcal{G})=SV \alpha$ depends on $\mathcal{G}$. Note that $f_1(\alpha,\overline{k},0)=f(\alpha,\overline{k})$ as defined previously.  

\begin{figure} 
\centering      \includegraphics[width=7.7cm]{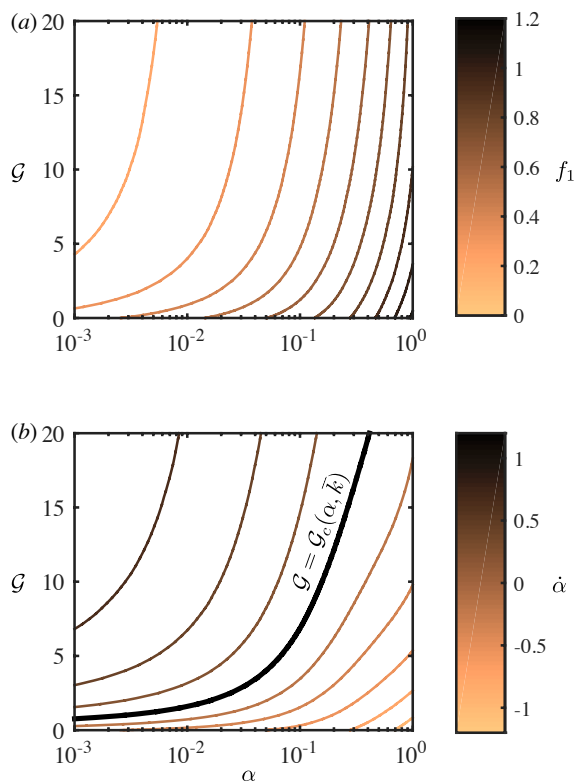}
\caption{Contours of:  (\textit{a}) the radial growth rate $f_1$ and (\textit{b}) the rate of change in aspect ratio $\dot{\alpha}$. In the latter, we highlight the critical curve $\mathcal{G}=\mathcal{G}_c(\alpha,\overline{k})$ on which the aspect ratio is constant ($\dot{\alpha}=0)$. Throughout we take $\overline{k}=4$. 
} \label{fig:axial_comb}
\end{figure}

Firstly, axial growth inhibits radial growth. Thus $f_1$ decreases as $\mathcal{G}$ increases across the whole parameter space (figure \ref{fig:axial_comb}\textit{a}). Axial growth releases latent heat, which increases the temperature of the top of the disk and so reduces conduction through the disk interior away from the radially growing edge of the disk. This is especially significant because the disk is a good thermal conductor and a significant fraction of the removal of latent heat for solidification at the disk edges occurs via transport through the solid disk interior. 

Secondly, the dimensionless axial growth increases with $\mathcal{G}$, and this effect is stronger at moderate aspect ratios  (\textit{cf}. figure \ref{fig:axial-laws}\textit{a}) because the faces are further from the melting temperature. 

\begin{figure} 
\centering      \includegraphics[width=8.0cm]{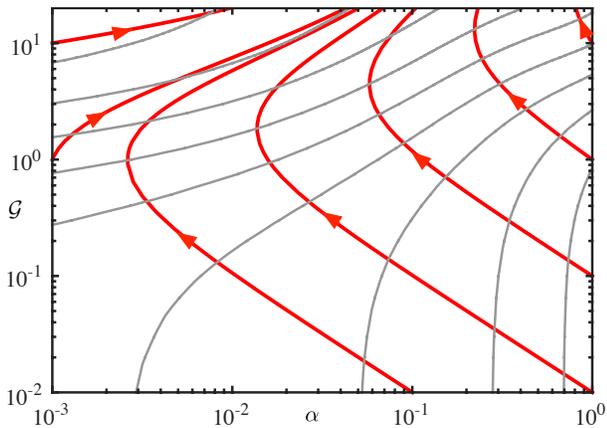}
\caption{Phase portrait showing trajectories in parameter space of crystal aspect ratio $\alpha$ and kinetic coefficient $\mathcal{G}$ when $\overline{k}=4$. The thick dark red curves with arrows show trajectories and the dashed light grey curves are contours of $\dot{\alpha}$. If the material properties of the crystal are held fixed, then variation in $\mathcal{G}$ directly corresponds to variation in the dimensional crystal radius $R$ to within a constant proportionality factor.} \label{fig:traj}
\end{figure}

Thirdly, we can compare axial and radial growth by considering the rate of change of aspect ratio. Using our quasi-steady predictions of heat transfer to predict instantaneous growth rates $V$ and $W$ for given values of $R$ and $H$, we derive a simple autonomous system for the dimensionless kinetic coefficient $\mathcal{G}$ and the aspect ratio $\alpha=H/R$,
\begin{align} 
\dot{\mathcal{G}}&=\mathcal{G} f_1/\alpha, \label{eq:autoG} \\
\dot{\alpha}&=f_2-f_1, \label{eq:autodoth} 
\end{align}
where a dot represents a derivative with respect to the slow timescale $\tau=t/\mathcal{S}$. 
In figure \ref{fig:axial_comb}(\textit{b}), we highlight the critical curve $\dot{\alpha}=0$, which, for aspect ratios $10^{-2}<\alpha<10^{-1}$ typical of frazil ice, corresponds in order of magnitude to $1<\mathcal{G}<10$. Below the critical curve the aspect ratio decreases, and so the crystal becomes more elongated. Conversely above the critical curve the aspect ratio increases. There is an important qualitative difference between $\mathcal{G}=0$ and $\mathcal{G}>0$. When $\mathcal{G}=0$, $\dot{\alpha}<0$ for all $\alpha$ (the thickness is fixed but the radius increases so the aspect ratio decreases). However, when $\mathcal{G}>0$, there is a critical aspect ratio below which $\dot{\alpha}>0$. This can be used to interpret crystal size evolution. Soon after a crystal nucleates, $\mathcal{G}$ will be small but the aspect ratio will be $O(1)$. As the crystal radius increases, the aspect ratio decreases towards the critical curve, but $\mathcal{G}$ will increase. Thus, at sufficiently late time, the aspect ratio will eventually start to increase. Some such trajectories are shown in figure \ref{fig:traj}. It is important to note that the timescale used in the non-dimensionalization is proportional to $R^2$ and so the evolution of a crystal in phase space slows down as the crystal radius $R$ increases. 

The autonomous system of equations (\ref{eq:autoG}--\ref{eq:autodoth}) significantly simplifies the parameter space. For example, we have scaled out the dependence on supercooling for the linear growth laws, which is hard to hold constant experimentally. Thus this is a potentially powerful way to interpret experimental data by plotting time series of experimental observations in this parameter space. Observing a minimum aspect ratio and the radius at this aspect ratio could be used to infer the dimensional kinetic coefficient $G$. We show a phase portrait of this autonomous system in figure \ref{fig:traj}.

Numerically, we observe that the critical curve, $\mathcal{G}=\mathcal{G}_c(\alpha,\overline{k})$, on which $\dot{\alpha}=0$,  approaches zero as $\alpha\rightarrow 0$. In the particular case $\overline{k}=1$, we can average  the solution (equation 7a  of  \citet{Fujioka74}) over the surface of the disk, to show that
\begin{equation}
\mathcal{G}_c\sim \frac{\pi}{\log (\alpha^{-1})-3(1-\log 2)}.
\end{equation}
Convergence is exponentially slow as $\alpha\rightarrow 0$ , but this nevertheless illustrates the important result that $\mathcal{G}_c\rightarrow 0$ as $\alpha \rightarrow 0$, which affects the range of possible trajectories in phase space.

\section{Combined heat and salt transfer} \label{sec:binary}
The bulk diffusion of salt can play a leading order role in growth from the liquid phase through two related physical mechanisms. Firstly, the presence of salt reduces the freezing temperature of the solution, resulting in a smaller imposed far-field supercooling for a given far-field temperature. Secondly, the ice crystal rejects salt during growth, which can build up locally at the interface, so further inhibiting growth by depressing the interfacial temperature. We here focus on their impacts on disk shaped crystal growth, noting that in certain circumstances these effects may also promote morphological instability. We return to the latter possibility in the concluding discussion. 

\subsection{Extended problem formulation}
We extend our method by additionally solving for the solute concentration field $C$ outside the disk, assuming that the concentration inside the disk $C_s$ is constant because diffusion of salt through the solid phase is slow relative to diffusion through the liquid phase. We outline the method more briefly. The main difference is that we require a condition relating the interfacial temperature $T_i$ to the concentration $C_i$ at the interface to couple the heat and salt problems. Thus on the growing edge we impose 
\begin{equation} \label{eq:Ti(C)}
T_i=T_L(C_i) \equiv T_m-m\left(C_i-C_s\right),
\end{equation}
where $m$ is the gradient of the (assumed linear) liquidus relationship $T_L(C)$ and $T_m=T_L(C_s)$ is the melting temperature of solid with concentration $C_s$. We assume  $C$ approaches a uniform concentration $C_\infty$ far from the crystal. 

We use a dimensionless concentration $\Theta=(C-C_\infty)/ \Delta C_\infty$, where $\Delta C_\infty=C_i-C_\infty$. Note that the non-dimensionalization involves $C_i$, which must be determined as part of the solution. In the coupled problem we must redefine $\theta=(T-T_\infty)/(T_i-T_\infty)$ and $\mathcal{S} =\rho_s L/\rho_l c_l \left(T_i - T_\infty \right)$, which depend on $\left(C_i-C_s\right)$ through the liquidus relationship~\eqref{eq:Ti(C)}. We define $\mathcal{C}=(C_i-C_s)/(C_i-C_\infty)$  which is the compositional analogue of the Stefan number and again must be large for the quasi-steady approximation to hold (section \ref{sec:QS}). Note that the thermal problem takes the same form as before, but with $\mathcal{S}=\mathcal{S}(\mathcal{C})$. 

Thus we must additionally solve $\nabla^2 \Theta=0$ outside the disk subject to the following boundary conditions, which are analogous to equations (\ref{eq:bctop}--\ref{eq:bcT}), 
\begin{equation}
 \left. \frac{\partial \Theta}{\partial z}\right|_{z=\alpha^+}=0,
\end{equation}
on the top of the disk $(0\leq r \leq1)$ where we set $G=0$, and
\begin{align}
&Le\mathcal{C} V =-  \left. \frac{\partial \Theta}{\partial r}\right|_{r=1^+},   \\
& \left.  \Theta \right|_{r=1^+} =1, \label{eq:BCTheta}
\end{align}
on the growing edge $(0\leq z\leq \alpha)$. As before, equation \eqref{eq:BCTheta}  is only applied at $z=0$. The Lewis number,
\begin{equation}
Le=\kappa_l / D_l,
\end{equation}
is the ratio of diffusivity of heat  $\kappa_l$ to solute $D_l$ in the liquid phase. 

We calculate a rescaled growth rate $g(\alpha)=Le \mathcal{C} V \alpha$ for the salt diffusion problem   (shown in figure \ref{fig:salt-deph}), and eliminate $V$ using~\eqref{eq:fdef} from the thermal problem to obtain
\begin{equation} \label{eq:Cquad}
Le \mathcal{C} = \frac{g(\alpha)}{f(\alpha,\overline{k})} \mathcal{S}(\mathcal{C}).
\end{equation}

\subsection{Results}
\begin{figure} 
\centering    \includegraphics[width=7.8cm]{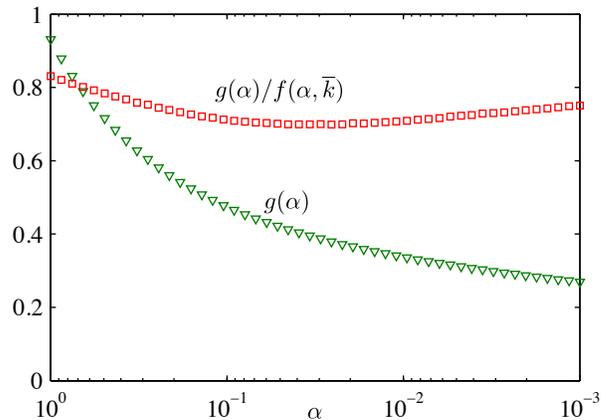}
\caption{The growth rate for salt diffusion problem $g(\alpha)$ (green triangles), and the combined problem $g(\alpha)/f(\alpha,\overline{k}=4)$ (red squares). We present approximate fits to these results in appendix \ref{app:formulae}.} \label{fig:salt-deph}
\end{figure}

There are three independent temperature scales in the problem
\begin{align*}
\Delta T_\infty &= T_m-T_\infty, \\
\Delta T_L&=\rho_s L / \rho_l c_l, \\
\Delta T_C&= m \left(C_\infty-C_s\right).
\end{align*}
The remaining parameters only appear in the group  $g(\alpha) / f(\alpha,\overline{k})Le$. In order to group separately what might be considered material and geometry-dependent parameters versus experimental parameters,  we define
\begin{equation} \label{eq:def-hatS}
\hat{\mathcal{S}}=\frac{\Delta T_L}{\Delta T_C} \frac{1}{Le} \frac{ g(\alpha)}{ f(\alpha,\overline{k})}, 
\end{equation}
representing the importance as the crystal grows of the diffusive removal of the latent heat released during crystal growth to the diffusive removal of rejected solute and the resulting freezing point depression. When $\overline{k}=4$, the ratio $g(\alpha) / f(\alpha,\overline{k})$ shows only weak variation with $\alpha$, as shown  in figure \ref{fig:salt-deph}. Thus $\hat{\mathcal{S}}$ could reasonably be treated as a material constant during disk growth, to leading order. 

We also define the dimensionless supercooling $\beta$ through 
\begin{equation}  \label{eq:def-beta}
\Delta T_\infty = \Delta T_C(1+\beta),
\end{equation}
where $\beta>0$ ensures supercooling in the far-field. Indeed,  there is supercooling everywhere in the liquid, and equilibrium is only achieved on the growing ice--liquid interface. (This follows by noting that the liquidus relation $T_L(C)$ is linear so  $\nabla^2[T-T_L(C)]=0$ and then applying the maximum principle for Laplace's equation.) Using equations~\eqref{eq:def-hatS} and~\eqref{eq:def-beta}, to express equation \eqref{eq:Cquad} in terms of $\mathcal{C}$ yields a quadratic equation with solution
\begin{equation}
\mathcal{C}=1+\frac{1+\hat{\mathcal{S}}-\beta + \sqrt{(1+\hat{\mathcal{S}}-\beta)^2+4\beta}}{2\beta},
\end{equation}
where we take the positive square root since $\mathcal{C}>1$ from the definition. Note that we actually require $\mathcal{C}\gg1$ for the quasi-stationary approximation to hold, consistent with weak supercooling $\beta \ll 1$.

To gain insight into the impact of solute on the growth rate, we investigate the factor $\mathcal{V}$ by which salt modifies crystal growth relative to growth into a pure melt with supercooling adjusted for the salt
\begin{equation}
\mathcal{V}=\frac{V}{(f(\alpha,\overline{k})/h) \cdot \left(T_L(C_\infty)-T_\infty\right) / \Delta T_L} =  \frac{\hat{\mathcal{S}} }{\beta \mathcal{C} }.
\end{equation}
In the physically relevant limit of small supercooling $\beta \rightarrow 0$,
 \begin{equation} \label{eq:Vratio-asym}
\mathcal{V} \sim  \frac{\hat{\mathcal{S}}}{1+\hat{\mathcal{S}}} -\beta  \frac{\hat{\mathcal{S}}}{(1+\hat{\mathcal{S}})^3}+\beta^2  \frac{\hat{\mathcal{S}}(1-\hat{\mathcal{S}})}{(1+\hat{\mathcal{S}})^5}+O(\beta^3) .
\end{equation}
We can subsequently take limits of the leading order term for small and large $\hat{\mathcal{S}}$: 
\begin{align} 
\mathcal{V} &\sim \hat{\mathcal{S}}  \quad (\hat{\mathcal{S}}\rightarrow 0),  \label{eq:Vratio-lima} \\   
\mathcal{V} &\sim 1 \quad (\hat{\mathcal{S}}\rightarrow \infty), \label{eq:Vratio-limb}
\end{align}
which we interpret in the discussion below. 

\section{Discussion and Conclusions} \label{sec:discussion}
We now apply the theoretical results from the preceding sections to infer the physical consequences for predictions of crystal growth, and evaluate some previous more approximate parameterisations. 
\subsection{Dimensional results for purely radial growth}
The purely radial growth rate of a disk-shaped crystal into a pure melt, in dimensional terms, is 
\begin{equation} \label{eq:dimV}
V_{\mathrm{dim.}} =\frac{1}{H} \frac{k_l \Delta T_\infty}{\rho_s L} f(\alpha,\overline{k}).
\end{equation}
For a binary alloy, we recover the pure melt case in the limit of large $\hat{\mathcal{S}}$ (equation \ref{eq:Vratio-limb}) with an adjusted driving temperature difference $\Delta T_e=T_L\left(C_\infty \right)-T_\infty$, so
\begin{equation} 
V_{\mathrm{dim.}} \sim \frac{1}{H} \frac{k_l \Delta T_e}{\rho_s L} f(\alpha,\overline{k}).
\end{equation}
This means that if the dimensionless group $\hat{\mathcal{S}}$ is sufficiently large, a good modelling assumption is to use formulae appropriate to a pure melt but adjust the freezing temperature when calculating supercooling to account for the solute impurity. Growth is controlled by the diffusive removal of the latent heat released during solidification. However, for small $\hat{\mathcal{S}}$ (equation \ref{eq:Vratio-lima}) we find
\begin{equation}
V_{\mathrm{dim.}}\sim \frac{1}{H} \frac{D_l \Delta T_e}{\Delta T_C} g(\alpha) = \frac{1}{H}D_l \beta g(\alpha),
\end{equation}
which means that growth is no longer controlled by the thermal diffusion of latent heat released at the interface but rather by the slow diffusion of solute rejected there.

\subsection{A simple way to account for the presence of salt}
Salt  significantly affects frazil ice growth in the ocean. To see this, we estimate typical values $\Delta T_L=80^\circ$C,   $\Delta T_C=2^\circ$C, and $Le=200$ to 1 significant figure,  using material properties estimated at $0^\circ$C, ocean water of salinity $C_{\infty}=35\,\mathrm{g\,kg}^{-1}$ and pure ice with $C_s\approx 0$. Thus $\hat{\mathcal{S}}\approx 0.16$ which is is an intermediate case with $\hat{\mathcal{S}} \lesssim O(1)$, but rather closer to the limit dominated by solute rejection. Thus both the dependence of freezing temperature on salinity and solute rejection are important, and significant errors result from neglecting either. There is large error in assuming that growth is controlled by the removal of released latent heat alone. 

In larger scale models that parameterize frazil-ice growth, it is very common to adjust the freezing temperature with salinity, and some models also investigate the effect of salt rejection and diffusion. For example, \citet{Holland05} multiply the growth rate by 0.2 as a way of testing for the sensitivity to salt. \citet{Galton-Fenzi12}, extending \citet{Holland99}, multiply their growth rate by a factor of $1/(1+Le \Delta T_C / \Delta T_L)$ which is typically about 0.2. Now at small supercooling, equation \eqref{eq:Vratio-asym} gives
\begin{equation} \label{eq:Vsim}
\mathcal{V} \sim  \left[ 1+ Le \frac{\Delta T_C}{\Delta T_L} \frac{f(\alpha,\overline{k})}{g(\alpha)}\right]^{-1},
\end{equation}
which is a similar expression since the ratio $g(\alpha)/f(\alpha,\overline{k})$ is of order 1. Therefore, the approach of \citet{Galton-Fenzi12} is likely to capture correctly the leading order behaviour for the salinity dependence of growth, although using equation \eqref{eq:Vsim} with the numerical dependence on aspect ratio from equation \eqref{eq:appgf} could give slightly better results.

\subsection{Growth rate of crystal mass and scaling analysis} \label{sec:scaling}
\begin{figure} 
\centering    \includegraphics[width=8.6cm]{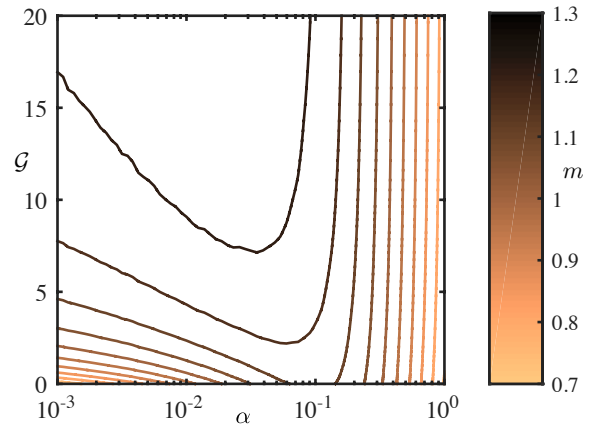}
\caption{The dimensionless crystal mass growth rate $m$ into a pure melt when  $\overline{k}=4$, relevant to water-ice systems.} \label{fig:contM}
\end{figure}

An instructive way to express disk crystal growth when the crystal grows both radially and axially is in terms of growth rate of crystal mass. We write
\begin{equation} \label{eq:mass}
L\frac{d M}{dt}=A \frac{k_l \Delta T_\infty}{R}   m(\alpha,\overline{k},G),
\end{equation}
where $M$ is the mass of the crystal, $A$ is the total surface area, and the effective total growth rate factor $m=(2 f_1 +f_2)/(1+2\alpha)=O(1)$, as shown in figure \ref{fig:contM}. These results do not depend significantly on the choice of axial growth law. As a crude simplification, it is possible to take $m = 1$ independent of $\alpha$ and $G$, so that the inclusion or exclusion of weak axial growth does not strongly impact the leading order scalings. If the aspect ratio is small, then $A\approx 2 \pi R^2$, so a simple formula for frazil-ice growth in a pure melt is 
\begin{equation} \label{eq:mass2}
L\frac{d M}{dt}\approx 2\pi R {k_l \Delta T_\infty},
\end{equation}
which can also be modified for salt as discussed previously. To within a factor of $4/\pi$, equation \eqref{eq:mass2} yields the same growth rate as equation (2) of \citet{Mason58}. 

Equation \eqref{eq:mass} has a simple physical interpretation in terms of the scaling arguments introduced in section \ref{sec:disc-intro}. Firstly, heat transfer occurs across the whole surface area $A$. For example, even when only the edges of the disk are growing, there is still a key contribution to the removal of latent heat from conduction through the solid from the growing edge and escaping through the crystal faces. Indeed the transfer through the faces dominates when $\overline{k} \alpha \gg 1$.  Secondly, the length scale of the thermal boundary layer scales with the radius of the crystal, not its thickness (see figure \ref{fig:EG-heat}), since $k_l \Delta T_\infty/R$ is a heat flux. These conclusions hold independently of the details of the axial growth.

In our opinion, the physical implications of the work of \citet{Fujioka74} have not been fully appreciated, because these scales could have been inferred from their mathematical model. We have extended their work to investigate the effect of different thermal conductivity ratios, an approximate description of axial growth, and salt. Many papers incorrectly estimate these scales controlling crystal growth. In some papers, for example \citep{Smedsrud04,Holland05,Holland07,Galton-Fenzi12,Hughes14}, the area for heat transfer $A$ is taken to be the area of the edge of the crystal $A\sim 4 \pi RH$ rather than the total surface area, whilst the thermal boundary layer is correctly assumed to have thickness proportional to $R$. The resulting heat transfer  thereby significantly underestimates crystal growth by a factor  between 10 and 100. Alternatively, in other earlier papers, for example \citep{Jenkins95}, the correct order of magnitude for the growth rate is obtained by erroneously using the area of the crystal edge $A\sim 4 \pi RH$ combined with a thermal boundary layer thickness proportional to $H$, two errors in the derivation that cancelled to produce the correct order of magnitude for the final result. Moreover, in models that use a distribution of crystal sizes following \citep{Smedsrud04}, our results suggest that there has been an underestimation of evolution towards the larger crystal sizes, an area of research we are actively pursuing.

\subsection{Implications and limitations} \label{sec:limitations}
We have identified order of magnitude errors in predictions of ice growth controlled primarily by thermal and solutal diffusion, or equivalently the timescale over which the initial supercooling of a melt is relieved. Turbulent heat transfer will play a role when the crystal radius is larger than the Kolmogorov length, because the thermal boundary layer scales like the crystal radius. While all of our analysis is confined to diffusive growth, much carries through relatively straightforwardly to the case of relatively weak turbulence by multiplying the diffusive growth rate by a modified heat transfer coefficient \citep{Batchelor80,Daly84}. It is therefore important to characterize correctly the diffusive growth of crystals, and our calculations have rationalized this process and allowed us to test the assumptions inherent in models of frazil-ice dynamics. Our calculated growth rates are likely to be important in more detailed models of frazil-ice population dynamics that account for evolution in crystal-size distribution \citep{Smedsrud04}. 

The assumption of small supercooling has entered this analysis at a number of stages. The quasi-steady approximation requires that the supercooling is small compared to $\Delta T_L\approx 80^\circ$C for growth from pure water, and, for the case of ocean water of salinity $C_{\infty}=35\,\mathrm{g\,kg}^{-1}$, the supercooling must be small compared to $\Delta T_c \approx 2^\circ$C. Small supercooling inhibits morphological instability since experimental observations and stability analyses suggest that disk-shaped crystals are morphologically stable provided the thickness is less than about 100 times greater than the nucleation length \citep{Shimada97,Yokoyama09}, and this length is inversely proportional to the supercooling. The supercooling observed in lakes and  oceans is typically rather small (for example, the largest supercooling recorded by \citet{Skogseth09} in an Arctic polynya was \mbox{$0.037^\circ$C}), and so our assumption to neglect morphological instability appears to be appropriate for frazil ice. For stronger supercooling, a range of crystal morphologies occur and a more complex study of heat transfer is required. 

The long range, diffusive transport of heat and salt plays an important role in the solidification of disk-shaped crystals from a binary melt. We have identified the physical scales that determine the bulk growth rate. We used a simple, thermodynamically motivated, anisotropic kinetic coefficient consistent with  a disk-shaped morphology, and neglected anisotropic surface energy. In doing so we provided a complementary perspective on crystal growth to that of `kinetic Wulff shapes' relevant to faceted snow ice growth, which allowed us to highlight the role of diffusion to the growth of disk-shaped frazil ice in oceans and rivers.

\begin{acknowledgments}
We would like to thank Jerome Neufeld and John Wettlaufer for valuable comments on earlier versions of this work. This publication arises from research funded by the John Fell Oxford University Press (OUP) Research Fund, and AJW also acknowledges financial support through the research program of the European Union FP7 award PCIG13-GA-2013-618610 SEA-ICE-CFD. 
\end{acknowledgments}

\appendix

\section{Numerical method} \label{app:numerics}
We adopt a straightforward numerical method. We solve the axisymmetric form of Laplace's equation in $(r,z)$ space using a Finite-Element-Method with adaptive meshing, which concentrates the mesh near the disk corner, where most resolution is needed. We used the MATLAB Partial Differential Equation Toolbox. We use linear basis functions on triangular elements. We truncate our domain at spherical radius $\tilde{r}=\tilde{R}$, following the method of \citet{Bayliss82}. Setting $\theta=0$ on this outer boundary gives an $O(1/\tilde{R})$ error. Thus, motivated by the well-known multipole expansion for far-field behaviour of the solutions of Laplace's equation, we instead set 
\begin{equation} \label{eq:theta-far}
\frac{\partial \theta}{\partial \tilde{r}}+\frac{\theta}{\tilde{r}}=0, \qquad (\tilde{r}=\tilde{R})
\end{equation}
which has an   $O(1/\tilde{R}^2)$ error.

In order to implement the jump boundary condition equation \eqref{eq:bcDTDR}, we introduce a notch of thickness $\epsilon$ at the growing edge of the disk, in which we impose a volumetric heat source. The notch becomes a line source in the limit $\epsilon \rightarrow 0$. We investigated the dependence of $f(\alpha,\overline{k})$ and hence the growth rate on $\epsilon$ and $\tilde{R}$. We ensure convergence to a relative error of  less than 0.2\% across  the entire  parameter space considered by using  $\tilde{R}=32$, $\epsilon=2 \times 10^{-7}$. Our results were benchmarked against the analytical solution of  \citet{Fujioka74} for the case $\overline{k}=1$, as illustrated in figure~\ref{fig:heat-comp}.

In terms of axial growth (section \ref{sec:axial}) the non-dimensional boundary conditions on the disk face are 
\begin{align}
\mathcal{S}W&=\left[ k \frac{\partial \theta}{\partial z} \right]_l^s, \label{eq:StefanW} \\
W&=\mathcal{G} \int_0^1 2 (1-\theta_i) r \, dr \label{eq:kineticW}
\end{align}
from equations \eqref{eq:Stefan} and \eqref{eq:kinetic} using the first axial growth law \eqref{eq:ax1} as an example.  We solve equation \eqref{eq:StefanW} in the same way as \eqref{eq:bcDTDR}, by introducing a notch. The crucial difference to the purely radial growth case is that equation \eqref{eq:kineticW} introduces a nonlinearity into the system of equations, which we solve iteratively, using a Newton-Raphson method.

\section{Practical formulae derived from fits to numerical calculations} \label{app:formulae}
Motivated by the asymptotic form of the toroidal integral (equation \ref{eq:toroidal}), we look for fits of the form $1/(b-c\log(x))$, for $b,c$ constant, to the numerically calculated results presented in figures \ref{fig:heat-comp} and \ref{fig:salt-deph}. We obtain
\begin{align}
f(\alpha,\overline{k}=1)&\approx 1/(0.9675-0.3160\log(\alpha)), \\
f(\alpha,\overline{k}=4)&\approx 1/(0.9008-0.2634\log(\alpha)), \\
g(\alpha)&\approx 1/(1.100-0.4146\log(\alpha)). 
\end{align}
The absolute errors in these formulae are typically very small, and are entirely negligible compared to the modelling uncertainties. Depending on the range of $\alpha$ of interest, different formula can be obtained, but these are practical for $10^{-3}< \alpha<1$.

For the ratio $g(\alpha)/f(\alpha,\overline{k}=4)$ important to the combined heat and salt calculation, a very accurate formula is
\begin{equation} \label{eq:appgf}
\frac{g(\alpha)}{f(\alpha,\overline{k}=4)}=\frac{0.9457 \alpha^2+2.775 \alpha+18.08}{\alpha^2+1.574 \alpha+21.79}.
\end{equation}
A simpler alternative with slightly diminished accuracy is to use a constant value  as mentioned in the main text, for example 0.73.

\end{document}